\documentclass[aps, prl, twocolumn, amsmath, graphicx, latexsym, asmsymb, amsfonts]{revtex4}

\usepackage{epsfig}

\DeclareMathOperator{\tr}{tr}

\begin{document}

\title{Capacities of Quantum Channels for Massive Bosons and Fermions}

\author{Aditi Sen(De)$^{1,2}$, 
Ujjwal Sen$^{1,2}$, 
 Bartosz Gromek\(^3\), Dagmar Bru{\ss}$^4$, and Maciej Lewenstein$^{1,2,*}$}

\affiliation{\(^1\)ICFO-Institut de Ci\`encies Fot\`oniques, 
Parc Mediterrani de la Tecnologia,
E-08860  Castelldefels (Barcelona), Spain\\
$^2$Institut f\"ur Theoretische Physik, 
Universit\"at Hannover, D-30167 Hannover,
Germany\\
\(^3\)Department of Theoretical Physics, University of Lodz, ul. Pomorska 149/153, PL-90236  Lodz, Poland\\
$^4$Institut f\"ur Theoretische Physik III, Heinrich-Heine-Universit\"at D\"usseldorf, 
D-40225 D\"usseldorf, Germany
}

\begin{abstract}

We consider the capacity of classical information transfer for noiseless quantum channels carrying a finite average number of 
massive bosons and fermions. 
The maximum capacity is attained by transferring the Fock states
generated from  the grand-canonical ensemble. 
Interestingly, the channel capacity for a Bose gas indicates the onset of a Bose-Einstein condensation, by 
changing its qualitative behavior at the criticality, while
for a channel carrying weakly attractive fermions, it exhibits the signatures  of  
Bardeen-Cooper-Schrieffer transition. We also show that for noninteracting particles, fermions are better carriers of information than 
bosons.

\end{abstract}
\maketitle

A communication channel carrying classical information by using quantum states as the 
carriers of information, has been a subject of intensive  studies. The fundamental 
result in this 
respect is the ``Holevo bound'' \cite{Holevo}  
(see also e.g. \cite{onno, Schumacher-er_ghyama, byapari-alada}), 
obtained more than 30 years ago, which gives the capacity of such channels.
An essential message carried by the Holevo bound is that \emph{at most \(n\) bits (binary digits) of 
classical information 
 can be sent via a quantum system of \(n\) 
distinguishable
qubits (two-dimensional quantum systems)}. 
However in realistic channels, where the quantum system is usually of infinite dimensions, 
the Holevo bound predicts  infinite capacities. In realistic channels, it is therefore important to
give a physical constraint on the carriers of the information.

Information carried over long distances usually employs electromagnetic signals as 
carriers of information. Capacities of such channels have been studied quite extensively (see e.g. 
\cite{rmp,onno, onno2}). 
In this case, the physical constraint that is used to avoid the infinite capacity 
problem, is an  energy constraint. Due to the form of the Holevo bound, the ensemble that 
maximizes the capacity,
turns out to be the canonical ensemble (or the microcanonical ensemble, depending on the type of the energy 
constraint) of statistical mechanics \cite{rmp}.

In recent experiments, it has been possible to produce atomic waveguides in optical microstructures \cite{hannover}, 
or on an atom chip \cite{revchip}, that may serve as  quantum channels of 
macroscopic (or at least mesoscopic) length scales.
Channels carrying massive particles have possibly fascinating  
applications in quantum information processing. It is thus important  to 
obtain their capacities.
Since we are dealing now with massive information carriers, it is 
not enough to put an energy constraint only. Rather, it is natural to give a particle number constraint as well as 
an energy constraint. This, of course, hints at the grand-canonical ensemble (GCE) of statistical mechanics. Indeed, we 
show in this Letter, the ensemble that maximizes the capacity of noiseless channels that carry massive bosons or fermions, 
under particle number and energy constraints, is GCE. 
Note that massless photons, despite being bosons, do not 
exhibit  Bose-Einstein condensation (BEC), due to the lack of constraint on their number. Massive bosons, however, do exhibit  BEC; 
we show in this Letter that the channel capacity of massive bosons
indicates the onset of BEC, by changing its behavior from being concave with respect 
to temperature, to being convex. The bosons that we consider in this Letter are noninteracting.  Noninteracting fermions, however, do not  
exhibit any phase transition. Interacting fermions, on the other hand,  exhibit the Bardeen-Cooper-Schrieffer (BCS) transition, and as we show 
in this Letter, the capacity
of interacting fermionic channels exhibits the onset of such transition. We obtain our results by simulating  a finite number of 
particles in the channel, and not the thermodynamical limit of an infinite number of particles.
We also show that for a wide range of power law potentials, including the harmonic trap and the rectangular box, and for moderate and high
temperatures, the fermions are better carriers of information than bosons, for the case of noninteracting particles. 

Suppose therefore that a sender (Alice) 
%
%
encodes the classical message \(i\) (occuring with probability \(p_i\)) in the state \(\varrho_i\), and sends it to a receiver (Bob). 
The channel is noiseless, while \(\varrho_i\) can be mixed.
To obtain  information about \(i\), Bob performs a measurement \(M\) (on the ensemble \({\cal E} = \{p_i, \varrho_i\}\)) 
to obtain the 
post-measurement ensemble  \(\{p_{i|m}, \varrho_{i|m}\}\),
with probability \(q_m\). 
The information gained by this measurement can be quantified by the mutual information \(I_M({\cal E})\)
between the index \(i\) and the measurement results \(m\): 
\(
I_M ({\cal E}) = H(\{p_i\}) - \sum_m q_m H(\{p_{i|m}\}).
\)
Here \(H(\{r_i\})= - \sum_i r_i \log_2 r_i \) is the Shannon entropy of a probability 
distribution \(\{r_i\}\). 
The accessible information
\(
I_{acc} ({\cal E}) = \max_M I_M ({\cal E})
\) is obtained by maximizing over all possible \(M\).

The Holevo bound gives a very useful upper bound on the accessible information for 
an arbitrary ensemble:
\(
I_{acc} ({\cal E}) \leq \chi({\cal E})   \equiv    S(\varrho) - \sum_i p_i S(\varrho_i)
\).
Here \(\varrho = \sum_i p_i \varrho_i\),
and \(S(\eta) = - \tr (\eta \log_2 \eta)\)
is the von Neumann entropy of \(\eta\). 
In a noiseless environment, the capacity of such an information transfer is the maximum, over all input ensembles
\emph{satisfying a given physical constraint}, of the accessible information.
%
%
It is important to impose a physical constraint on the input ensembles,
as
arbitrary encoding and 
decoding schemes are included in the Holevo bound. This has the consequence that the bound explodes for 
infinite dimensional systems: For an ensemble of pure states with average ensemble state \(\varrho_\psi\), 
\(\chi = S(\varrho_\psi)\), which can be as large as \(\log_2 d\), where \(d\) is the 
dimension of the Hilbert space to which the ensemble belongs. 
If
the pure states are 
orthogonal, \(I_{acc} = \chi = \log_2 d\), so that the capacity 
diverges along with its bound (see e.g. \cite{onno}).

To avoid this infinite capacity, one usually uses an energy constraint for channels \emph{that 
carry photons} (see e.g. \cite{rmp, onno, onno2}). Suppose that the system is described by the Hamiltonian \({\cal H}\).
Then the average energy constraint on a communication channel that is sending the 
ensemble \({\cal E} = \{p_i, \varrho_i\}\), is 
\(\tr(\varrho {\cal H}) = E\).
Here \(\varrho\) is the average ensemble state, and \(E\) is the average energy available to the system. 
The capacity \({\cal C}_E\) of such a channel is then the maximum of  
\(I_{acc} ({\cal E})\), over all ensembles, under the 
average energy constraint.
Now 
\(
I_{acc}({\cal E}) \leq \chi({\cal E}) \leq S(\varrho),
\)
and \(S(\varrho)\) is maximized, under the same constraint,
by the canonical ensemble (CE) corresponding to the Hamiltonian \({ \cal H}\), and energy \(E\) (see e.g. \cite{Huang}). 
Moreover, this 
is an ensemble of orthogonal pure states (the Fock states, or in other words number,  states), so that 
 \({\cal C}_E\) is also reached for this ensemble \cite{rmp}.
The channel capacity depends solely on the average ensemble state, which in this optimal case is 
the canonical equilibrium (thermal) state 
\(
\varrho^{eq} = \exp(- \beta {\cal H})/Z,
\)
where \(\beta = 1/k_B T\), with \(k_B\) being  the Boltzmann constant, and \(T\) the absolute temperature. 
\(Z = \tr(\exp(- \beta {\cal H}))\) is the partition function. 
For 
given 
\(E\), 
\(T\) is given by 
\(
E = - \frac{\partial}{\partial \beta} (\log_e Z).
\)
As a result of particle number nonconservation,
noninteracting photons and hence their
capacity do not exhibit signatures of a condensation. 
However, effectively interacting photon fluids and photon condensation effects are possible, in principle,
by using nonlinear cavities, in which case the photons may acquire an effective mass (see e.g. \cite{alor-jol}).


In the case of channels \emph{that carry massive particles}, it is natural to impose the \emph{additional} constraint 
of average particle number.  Suppose that Alice prepares  \(N\) particles in a trap, and transfers them to Bob. Let the trap have energy levels
\(\varepsilon_i\), and  let \(n_i\) be the average occupation number of the \(i\)-th level. 
Then the conservation of the average particle number reads
\(\sum_i n_i = N\),
and the constraint of a fixed average energy,
for a given energy \(E\), is 
\(\sum_i n_i \varepsilon_i = E\).
The channel capacity \({\cal C}_{E,N}\) of such a channel is the maximum of \(I_{acc}({\cal E})\), over 
all ensembles that satisfy these two constraints.
Under these constraints, the von Neumann entropy of the average state of the system is 
maximized by GCE. Again the ensemble elements are pure and orthogonal (Fock 
states), whence the channel capacity is reached by the same ensemble.

It is important to stress here that the channel capacities that we derive in this Letter are all 
for the case of a given \emph{finite} average number of particles in the trap, and not in the 
thermodynamic limit. This is because in a real implementation of such channels, this number 
is usually only at most moderately high. 

\emph{Noninteracting bosons.}  
%
%
Here, 
\(n^b_i = 1/(\mbox{e}^{\beta (\varepsilon_i - \mu^b)} -1)\), 
and the channel capacity (in bits) is given by (see e.g. \cite{Reichl})
\({\cal C}_{E,N}^{be} = - \sum_i \left( n^b_i \log_2 n^b_i - (1 + n^b_i) \log_2 (1 + n^b_i)\right)\).
Here \(\mu^b\) is the chemical potential.  For given average particle number  \(N\)
and absolute temperature \(T\), one uses 
the  energy constraint
to find \(\mu^b\) for that case. 
The 
energy is then given by 
the average particle number constraint,
and the capacity by \({\cal C}_{E,N}^{be}\).
Let us consider the case when the 
trap is a 3D-box of volume \(L^3\) \emph{with} periodic boundary condition (pbc), so that the energy 
levels
are \(\frac{2 \pi^2 \hbar^2}{mL^2} (n_x^2 + n_y^2 + n_z^2), \quad  n_x, n_y, n_z = 0, \pm 1,  \ldots
\).
Here \(m\) is the mass of the individual particles 
in the trap.

\begin{figure}[tbp]
\begin{center}
\epsfig{figure= 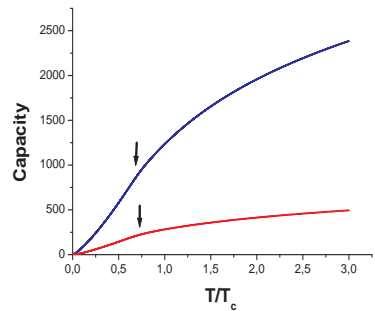,height=.2\textheight,width=0.3\textwidth}
\caption{
The channel capacity, for the case of noninteracting bosons, plotted against $T/T_c$. 
The lower curve is for the case of 100 bosons, in a 3D box
with pbc.
For 
$^{87}$Rb atoms in such a box of volume \(1 \mu\mbox{m}^3\), \emph{thermodynamical} calculations
predict \(T_c \approx 0.4 \mu\mbox{K}\). 
The upper curve is for 500 bosons in the same trap. 
The thermodynamical estimations of \(T_c\) are higher than the corresponding values that we obtain (as indicated by the arrows).   
}
\label{fig_boson_3d_1}
 \end{center}
\end{figure}

In Fig. \ref{fig_boson_3d_1}, 
we plot 
\({\cal C}^{be}_{E,N}\) vs.
\(T/T_c\), 
for 
different \(N\).
Here 
 \(T_c = (2 \pi \hbar^2/mk_B)(N/2.612L^3)^{2/3}\) is the critical temperature, 
 as obtained in the \emph{thermodynamical} (large \(N\)) limit. 
The 
capacity changes
its shape from being concave 
to being convex with respect to temperature, 
at the onset of a BEC. 
The \emph{thermodynamical} estimation of \(T_c\) is higher than our values of \(T_c\) for different \(N\), with the 
gap reducing for growing \(N\). Such indication of a gap has also been obtained previously (see e.g. \cite{Dalfovo}). 
We have checked that the predicted approximate gap in Ref. \cite{Dalfovo}, is in agreement with our calculations for 
\(N = 1000\). For lower \(N\) however, the prediction is no longer valid,  as expected in Ref. \cite{Dalfovo}. 
Note that the capacity increases with increasing \(N\),
%
%
%
%
%
%
%
%
and it 
has the same qualitative behavior for a 3D box \emph{without} pbc,
as well as for a harmonic trap.

\emph{Fermions are better carriers of information than bosons.} 
For spin-\(s\) noninteracting fermions, 
\(n^f_i = g/(\mbox{e}^{\beta (\varepsilon_i - \mu^f)} +1)\),
where \(g = 2 s + 1\), and \(\mu^f\) is the fermion chemical potential. 
The channel capacity (in bits) is given by (see e.g. \cite{Reichl})
\({\cal C}_{E,N}^{fd} = - g \sum_i \left( \frac{n^f_i}{g} \log_2 \frac{n^f_i}{g} + 
(1 - \frac{n^f_i}{g}) \log_2 (1 - \frac{n^f_i}{g})\right)\).
Again, for given \(N\) and \(T\), one obtains \(\mu^f\) from average particle number conservation,
which then gives the capacity.


The bosons that we have considered in this Letter are spinless. To make 
a fair comparison of the capacities, we consider ``spinless", i.e. polarized fermions with \(g=1\). 
Let us start with bosons, and perform the high temperature expansion. 
First, we expand the fugacity \(z^b=\mbox{e}^{\beta \mu^b}\) in powers of \(N/S_1\), where 
\(S_k = \sum_i \mbox{e}^{- k \beta \varepsilon_i}\), and find the coefficients from the average particle number conservation. 
We use this expansion to find an expansion of the \(n_i^b\)'s, which in turn is substituted in the formula for \({\cal C}^{be}_{E,N}\). 
The same calculation is done for fermions. We perform the calculation up to the third order, and find that 
\({\cal C}^{be}_{E,N} = (\sum_{i=1}^{3}\alpha^b_i (N/S_1)^i)\log_2\mbox{e} + \beta^b_1(N/S_1)\log_2(N/S_1) + \beta^b_2(N/S_1)^2\log_2(N/S_1)\), 
whereas 
\({\cal C}^{fd}_{E,N} = (\sum_i\alpha^f_i (N/S_1)^i)\log_2\mbox{e} + \beta^f_1(N/S_1)\log_2(N/S_1) + \beta^f_2(N/S_1)^2\log_2(N/S_1)\).
The coefficients of first order perturbation are equal: \(\alpha_1^b = \alpha_1^f = S_1 + D_1\). 
In the next order, they differ by a sign: \(\alpha_2^b = - \alpha_2^f = S_2/2 - S_2D_1/S_1 + D_2\). 
The third order perturbation coefficients are again equal: \(\alpha_3^b = \alpha_3^f = - 3S_2 + S_3/3 + 2S_2^2/S_1 + (2S_2^2 - S_1 S_3)D_1/S_1^2
- 2S_2D_2/S_1 + D_3\). Also, 
\(\beta_1^b = \beta_1^f = - S_1\), \(\beta_2^b =0\), \(\beta_2^f = 2S_2\);
here \(D_k = \sum_i \beta \varepsilon_i \mbox{e}^{- k \beta \varepsilon_i}\).

To find out the potentials and dimensions for which 
this perturbation technique is systematic, we consider uniform power law potentials, such as 
\(V = r^\gamma\), in a 
\(d\)-dimensional Cartesian space of \((x_1, \ldots, x_d)\), with \(r= \sqrt{x_1^2 + \ldots + x_d^2}\).
We calculate \(S_k\) and \(D_k\), replacing sums by 
integrations with density of states
\(\rho(\varepsilon) = \int d^d p d^d x \delta (p^2/2m + V - \varepsilon)\),  the latter integration 
being over the phase space. 
One may then check that the technique is systematic when \(1/\gamma + 1/2 > 1/d\). This includes, e.g., the 3D and 2D harmonic potentials,
for which we also performed the summations directly, and obtained the same results. 
Note that \(0=\beta_2^b < \beta_2^f\),  
whereas
\(\alpha_2^b= S_2(1 - \beta\left\langle \varepsilon \right\rangle_\beta + \beta\left\langle \varepsilon \right\rangle_{2\beta} )\)
(where 
\(\left\langle \varepsilon \right\rangle_\beta\) stands for the average energy with Boltzmann probabilities \(\mbox{e}^{-\beta\varepsilon_i}/S_1\))
can be explicitly evaluated by using the density of states \(\rho(\varepsilon)\).
As a result, we obtain \(\alpha_2^b \leq 0 \leq \alpha_2^f\), implying that 
\({\cal C}^{be}_{E,N} < {\cal C}^{fd}_{E,N}\) for a large range of sufficiently high temperature,  and for power law potentials
that satisfy the same condition as systematicity.
%
We have

\textbf{Theorem.} 
\emph{For power law potential traps (with power \(\gamma\) and dimension \(d\)), 
and for sufficiently high temperature, the capacity of fermions is better than that of bosons when 
\(1/\gamma + 1/2 > 1/d\).} 

\begin{figure}[tbp]
\begin{center}
\epsfig{figure= 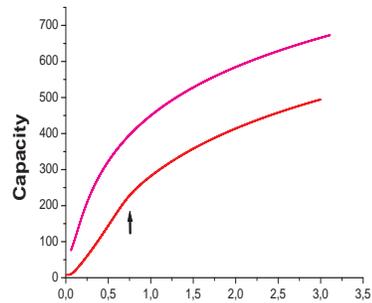,height=.2\textheight,width=0.3\textwidth}
\caption{
We compare the channel capacity for the case of 100 (noninteracting) spinless fermions (upper curve) with that of 100 
bosons (lower curve).  The trap is a 3D box with pbc.
For the lower curve, the horizontal axis is \(T/T_c\), while
for the upper curve, it is \(T/T_f\), 
where \(T_f = (\hbar^2/2mk_B)(6 \pi^2 N/gL^3)^{2/3}\) is the Fermi temperature. 
}
\label{fig_fermion_3d_1}
 \end{center}
\end{figure}

This includes, e.g., the harmonic trap in 2D and 3D, the 
3D rectangular box, and the 3D spherical box. The theorem holds for quite moderate \(T\), since we work up to the 
order \((N/S_1)^3\). 
Numerical simulations show indeed that it holds also for low temperatures, as seen, e.g., in 
%
%
%
%
%
%
%
%
%
Fig. \ref{fig_fermion_3d_1}
for 100 spinless fermions and same number of  spinless bosons
trapped in a 
3D box with pbc.
%

Note
that we have found that  
the capacities for a 3D box without pbc
are lower than those with pbc,
both for bosons and 
fermions. The capacity for a harmonic trap, with the same 
characteristic length scale, has a higher capacity (both for bosons  as well as for fermions) than that for a 3D box with pbc. 
More 
importantly, the capacities for the case of fermions do not show any signatures of criticality, as expected.  



\begin{figure}[tbp]
\begin{center}
\epsfig{figure= 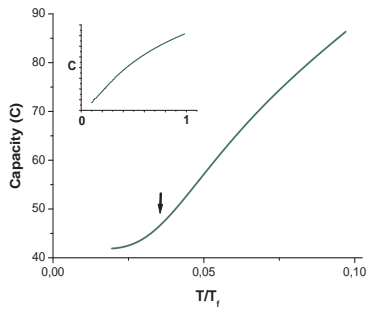,height=.2\textheight,width=0.3\textwidth}
\caption{
Channel capacity of 100 interacting fermions in a 3D box trap with pbc,
against $T/T_f$. The capacity is initially convex, and then becomes concave, for 
higher $T$'s, as illustrated in the inset. 
}
\label{fig_bcs}
 \end{center}
\end{figure}

\emph{Interacting fermions.} Until now, we have been dealing with the case of noninteracting bosons and fermions. 
Although a system of noninteracting massive bosons exhibits condensation, this is not
 the case for noninteracting fermions.
A system of interacting fermions, however,  \emph{can} exhibit Cooper pairing, and, consequently,  
superfluid BCS transition (see e.g. \cite{Reichl}).
It is therefore interesting to 
see whether such a ``condensation" can be observed in the capacity of a channel transmitting trapped 
interacting fermions.  We consider here a 3D box (of volume \(L^3\)) with pbc,
within which \(N\) fermions are trapped. The fermions behave like an ideal Fermi-Dirac gas, except when
pairs of them with equal and opposite momentum, and opposite spin components 
have kinetic energy 
within an interval \(\Delta \epsilon\) on either side of the Fermi surface.  
In that case, the pairs experience  a weak attraction. 
The Hamiltonian can then be written as 
\(
{\cal H} = \sum_{\vec{k}} t_{\vec{k}} (a^\dagger_{\vec{k},\uparrow} a_{\vec{k},\uparrow} 
                                                + a^\dagger_{\vec{k},\downarrow} a_{\vec{k},\downarrow}) 
+ \sum_{\vec{k}, \vec{l}} V_{\vec{k}, \vec{l}} 
a^\dagger_{\vec{k},\uparrow} a^\dagger_{-\vec{k},\downarrow}  a_{-\vec{l},\downarrow} a_{\vec{l},\uparrow}\),
where $t_{\vec k}=\hbar^2 k^2/2 m$, whereas 
\(V_{\vec{k}, \vec{l}}\) vanishes except when $ |\mu - \hbar^2 k^2/2m| \leq \Delta \epsilon$
and $ |\mu - \hbar^2 l^2/2m| \leq \Delta \epsilon$, in which case, 
\(V_{\vec{k}, \vec{l}} = - V_0 < 0\). 
Using mean field approximation, the average occupation number in this case turns out to be
$
N_{\vec{k}} = (1 - (\epsilon_{\vec{k}} / E_{\vec{k}}) \tanh (\beta E_{\vec{k}}/2))/2,
$
where 
\(\epsilon_{\vec{k}} =t_{\vec k}  - \mu\), \(E_{\vec{k}} = \sqrt{\epsilon_{\vec{k}}^2 + |\Delta_{\vec{k}}|^2}\).
Here \(\Delta_{\vec{k}} = \Delta(T)\), when \(|\epsilon_{\vec{k}}| \leq \Delta \epsilon\), and vanishing otherwise. 
This is the so-called ``gap'', given by the  equation, 
\(
\Delta_{\vec{l}} = V_0 \sum_{\vec{k}} (\Delta_{\vec{k}} /E_{\vec{k}}) \tanh(\beta E_{\vec{k}}/2),
\) 
where the summation runs only for an energy interval \(\Delta \epsilon\) about the Fermi surface. 
Note that the occupation numbers in this case are different from
the case of ideal (noninteracting) fermions. Using the gap equation and the constraint \(\sum_{\vec{k}} N_{\vec{k}} = N\)
on the total number of fermions in the trap, we can find the channel capacity
by replacing \(n^f_i\) by 
\(N_{\vec{k}}\) in \({\cal C}_{E,N}^{fd}\).  
As we see in Fig. \ref{fig_bcs}, the channel capacity again changes its behavior qualitatively, indicating the 
onset of the superfluid  BCS transition. In Fig. \ref{fig_bcs}, we plot the channel capacity against 
\(T/T_f\), where \(T_f\) is the Fermi temperature for the case of noninteracting fermions. This is for convenience, as 
the thermodynamical transition temperature (for interacting fermions) 
has an exponentially decaying factor, which renders it inconvenient for our purposes. Also we chose 
\(\Delta \epsilon = \frac{\hbar^2}{mL^2}\) and \(V_0 = 10^{-6} \frac{\hbar^2}{mL^2}\) in the figure.  

In this Letter, we have used the GCE, which has appeared due to 
the dual constraints of average energy and average particle number. 
For a fixed number of particles, and retaining 
the average energy constraint, we are led to CE. In the thermodynamical limit,
the average occupation numbers are the same for CE and GCE.
For finite \(N\),
exact calculations for CE
are difficult.
However, different approximate methods 
(see e.g. \cite{Nobel-kajer-por})
reveal that 
the average occupation numbers of CE
are more uniform as compared to GCE,
so that the former ensemble has
larger capacity. However, 
the difference is marginal. 

The channels that we have considered in this Letter are noiseless.
A simple, but physically important, 
model of noise is the Gaussian noise acting similarly on each 
mode, resulting in an 
effective increase of temperature in the channel.
So for a given temperature, 
to accomodate the average energy constraint, we must start 
with a lower temperature than that in the noiseless case, leading to a decrease in capacity. 
The lower capacity in this particular noisy case can be read off 
from the figures of the noiseless one after finding the temperature difference.

To conclude, we have considered the classical capacities of noiseless 
quantum channels carrying a finite average number of massive bosons or fermions. 
We have shown that the capacities are attained on the grand-canonical ensemble of statistical mechanics. 
Capacity of a channel 
carrying bosons indicates the onset of  Bose-Einstein condensation, by changing its behavior from being 
concave to convex 
with respect to the temperature, at the transition point. 
Also the signature of the onset of 
Bardeen-Cooper-Schrieffer transition
can be observed for 
weakly interacting fermions.
%
%
We show analytically that for noninteracting particles, fermionic channels are better than the bosonic ones, in a wide variety of cases.

We acknowledge support from the DFG
(SFB 407, SPP 1078, SPP 1116, 436POL), 
the Alexander von Humboldt Foundation, the Spanish MEC grant FIS-2005-04627,
the ESF Program QUDEDIS, and EU IP SCALA.

\end{document}